\documentclass[fleqn,usenatbib]{mnras}

\usepackage{newtxtext,newtxmath}

\usepackage[T1]{fontenc}
\usepackage{ae,aecompl}

\usepackage{graphicx}	
\usepackage{amsmath}	
\usepackage{amssymb}	

\title[inner Milky Way]{Heavy element evolution in the inner regions of the Milky Way}

\author[F. Matteucci et al.]{
F. Matteucci,$^{1,2,3}$\thanks{E-mail: matteucci@oats.inaf.it}
A. Vasini,$^{1}$
V. Grisoni$^{1,2}$
and M. Schultheis$^{4}$
\\
$^{1}$Department of Physics, Trieste University, Via G.B. Tiepolo, 11, 34131 Trieste, Italy\\
$^{2}$Italian National Istitute for Astrophysics (INAF), Via G.B. Tiepolo, 11, 34131 Trieste, Italy\\
$^{3}$Italian National Institute of Nuclear Physics (INFN), Via Valerio 2, 34100 Trieste\\
$^{4}$ Universit\'e Cote d'Azur, Observatoire de la Cote d'Azur, CNRS, Laboratoire Lagrange, France \\
}

\date{Accepted XXX. Received YYY; in original form ZZZ}

\pubyear{2015}

\begin{document}
\label{firstpage}
\pagerange{\pageref{firstpage}--\pageref{lastpage}}
\maketitle

\begin{abstract}
We present results for the evolution of the abundances of heavy elements (O, Mg, Al, Si, K, Ca, Cr, Mn, Ni and Fe) in the inner Galactic regions ($R_{GC}\le 4$kpc). We adopt a detailed chemical evolution model already tested for the Galactic bulge and compare the results with APOGEE data.  We start with a set of yields from the literature which are considered the best to reproduce the abundance patterns in the solar vicinity. We find that in general the predicted trends nicely reproduce the data but in some cases either the trend or the absolute values of the predicted abundances need to be corrected, even by large factors, in order to reach the best agreement. We suggest how the current stellar yields should be modified to reproduce the data and we discuss whether such corrections are reasonable in the light of the current knowledge of stellar nucleosynthesis. However, we also critically discuss the observations. Our results suggest that Si, Ca, Cr and Ni  are the elements for which the required corrections are the smallest, while for Mg and Al moderate modifications are necessary. On the other hand, O and K need the largest corrections to reproduce the observed patterns, a conclusion already reached for solar vicinity abundance patterns, with the exception of oxygen. For Mn we apply corrections already suggested in previous works.
\end{abstract}




\section{Introduction}

The project APOGEE (Apache Point Observatory Galactic Evolution Experiment, Majewski et al. 2017) belongs to the survey SDSS III/IV, and it has been devoted to observe a large sample of Galactic stars (roughly $10^{5}$) at the infrared wavelengths to obtain chemical and kinematical information. The choice of the infrared range allows us to observe regions obscured by dust such as the Galactic inner regions including the bulge.
These data offer a good opportunity to study the chemical evolution of the
inner Galactic regions as well as the timescale on which the majority of their stars formed. In fact, by comparing the data with predictions from chemical evolutionary models we can impose strong constraints on the formation and evolution of the inner Milky Way regions as well as on stellar nucleosynthesis. The main ingredients to build a chemical evolution model are: i) the stellar birthrate function, ii) the stellar yields and iii) the gas flows in and out. All of these ingredients contain uncertainties and especially important are the uncertainties in stellar yields. In fact,  C{\^o}t{\'e} et al. (2017) have investigated the impact of the different assumptions in chemical evolution models and concluded that successes and failures of such models are mainly due to uncertainties in the stellar yields rather than to the complexity of the galaxy model itself.\\
Previous models for the chemical evolution of the bulge (Matteucci \& Brocato, 1990; Ballero et al. 2007; Cescutti \& Matteucci, 2011; Grieco et al. 2012; Matteucci et al. 2019) have suggested that the majority of bulge stars formed out of a strong burst of star formation occurred on a relatively short timescale, no longer than 1 Gyr. According to the time-delay model (Matteucci, 2012) the predicted [$\alpha$/Fe] ratios in a regime of starburst appear larger than solar for a large range of metallicities, a fact predicted for the first time by Matteucci \& Brocato (1990). \\
The first observation of [$\alpha$/Fe] ratios in bulge stars was from Mc William \& Rich (1994) who confirmed a longer plateau for the [Mg/Fe] ratios in bulge stars, followed by  many other studies, such as Zoccali et al. (2003;2006), Cunha \& Smith (2006), Fulbright et al. (2007), Lecureur et al. (2007), Clarkson et al. (2008), Alves-Brito et al. (2010), Bensby et al. (2013), Johnson et al. (2014), Gonzalez et al. (2015), Bensby et al. (2017) and Rojas-Arriagada et al. (2017). In general, data on [$\alpha$/Fe] ratios have shown a slightly longer plateau than in the solar neighbourhood, although it is difficult to define the exact value of the [Fe/H] value at which the ``knee'' (change of slope) occurs, because of the spread in the data and the different results obtained by different knee indicators.\\
It is therefore quite important to test these hypotheses on the most recent data on the stars for the innermost regions of the Milky Way. In particular, we adopt as a reference the APOGEE data of Zasowski et al. (2019) and we consider the following chemical elements: O, Mg, Al, Si, Ca, K, Cr, Mn, Ni and Fe.  We exclude the elements Na and Co since the scatter in the data is too large and it does not allow us to draw any conclusion.
As stated above, the stellar yields, namely stellar nucleosynthesis, are very important ingredients in models of chemical evolution. Unfortunately, still many uncertainties are present in the stellar yields, as thoroughly discussed in Romano et al. (2010) and Prantzos et al. (2018). Because of such uncertainties, Fran\c cois et al. (2004) made the experiment of changing ad hoc the yields of chemical elements in order to perfectly match the observations. They started from the yields of Woosley \& Weaver (1995) for massive stars and those of Iwamoto et al. (1999) for Type Ia SNe and the elements considered were: O, Mg, Si, Ca, K, Ti, Sc, Ni, Mn, Co, Fe and Zn.
They suggested variations on the yields of those elements in order to fit the relations [X/Fe] vs. [Fe/H] in the solar vicinity. They pointed out large uncertainties for Fe-peak elements, which are related to the uncertain mass-cut applied to the nuclei of massive stars exploding as core-collapse SNe (CC-SNe), as well as for some $\alpha$-elements such as Mg, whose amount depends upon the uncertain rate of the $^{12}$C($\alpha, \gamma$)$^{16}$O reaction. For K the situation is complicated by the contribution to this element by neutrino-induced reactions. Those suggestions were thought to be helpful to nucleosynthesis modelers.\\
Later on, Romano et al. (2010) adopted different sets of literature yields and compared the results with solar vicinity data and concluded that the best set of yields includes the results of Kobayashi et al. (2006) for massive stars, except for C, N and O for which the yields of the Geneva Group were to be preferred, the yields of Karakas (2010) for low and intermediate mass stars and those of Iwamoto et al. (1999) for SNe Ia. Still, the abundances of several elements could not be reproduced, especially those of Fe-peak elements. They pointed out that several physical processes in stellar evolution  should be included and/or revised. These processes are the hot-bottom burning in low and intermediate mass stars, rotation in stars of all masses and mass loss in massive stars.
Here, we perform an exercise similar to that of Fran\c cois et al. (2004) by comparing the results of a model for the Galactic inner regions (Matteucci et al. 2019), adopting the best yields of Romano et al. (2010), with the infrared data of Zasowski et al. (2019). We find the variations that should be applied to the reference yields of the studied elements in order to obtain a very good fit to the observations. By doing that we achieve two goals: i) find a model that can reproduce a large number of chemical abundances in the inner Galactic regions and impose constraints on its formation and evolution history, and ii) impose constraints on the stellar nucleosynthesis.
The paper is organized as follows: in Section 2 we describe the stellar data, in Section 3 we present the chemical evolution model, in Section 4 we discuss our results. Finally, in Section 5 some conclusions are drawn.

\section{Observations}

In Zasowski et al. (2019) paper, the abundances of 12 chemical species (O, Na, Mg, Al, Si, K, Ca,  Cr, Mn, Co, Fe, Ni) are measured with the APOGEE Stellar Parameter and Chemical Abundances Pipeline (ASPCAP, Garcia-Perez et al. 2016) (see  also Majewski et al. 2017;  Nidever et al. 2015; Zasowski et al. 2013) in stars belonging to the inner Galactic regions ($R_{GC} \le 4$ kpc). The original APOGEE calibrated DR14/DR15 release (Abolfathi et al. 2018) includes 22 elements but they have been reduced to 12 by eliminating those elements for which the scatter is large (see J\"onsson et al., 2018). The spectra are taken in the infrared  wavelength range (between 1.51 and 1.69 microns) so to avoid dust absorption. The sample represents the data release 14/15 (DR14/DR15) and contains $\sim$ 4000 stars, selected in order to avoid surface gravities and metallicities not reliable and to have a homogeneous sample with surface temperatures in the range 3600-4500K, and surface gravities in the range from  log (g)= -0.75 to
+3.5. The quoted errors on the derived abundances are $<$0.1 dex. In the Zasowski et al. (2019) paper the abundance patterns [X/Fe] vs. [Fe/H] are described and compared with previous studies in the literature. For example, Johnson et al. (2014) measured the abundances of several chemical elements in 156 red giant stars in two Galactic bulge fields centered near (l,b)= (+5.25, -3.02) and (0, -12). The field (+5.25,-3.02) contains also observations of the bulge globular cluster NGC6553. The data originate from high resolution ($R\sim 20.000$), high signal to noise (S/N$ > \sim$ 70) spectra, obtained with FLAMES-GIRAFFE and belonging to European Southern Observatory archive. From these data, Johnson et al. (2014) selected the spectra that did not show strong TiO absorption lines.

\section{The model}
\subsection{Main assumptions}

The chemical evolution model for the Galactic bulge that we consider here is the one described in Matteucci et al. (2019). In that paper, we run a model for typical bulge stars  plus other models to reproduce a second population arising either from the inner disk or after a stop in the star formation during the bulge formation, and visible in the metallicity distribution function (MDF), which shows two peaks. Here, the stars we compare with belong to a larger region than the bulge and we adopt a unique model which is the basic one aimed at reproducing the majority of bulge stars.  
In this model the bulge forms by fast gas infall, with a timescale $\tau$= 0.1 Gyr. The model is one-zone and the assumed gas accretion law has an exponential form:

\begin{equation} \label{eq_01}
\dot G_i(t)_{inf}=A(X_i)_{inf}e^{-\frac{t}{\tau}},
\end{equation}
where $G_i(t)_{inf}$ is the infalling material in the form of the element $i$ and $(X_i)_{inf}$ the composition of the infalling gas which is assumed to be primordial.
The quantity $A$ is a parameter fixed by reproducing the present-time total surface mass density in the considered Galactic region, here the inner 4 kpc.\\
The star formation rate (SFR) is parametrized according to the Schmidt-Kennicutt law (Kennicutt 1998):
\begin{equation} \label{eq_03}
\text{SFR}(t)=\nu \sigma_{gas}^k(t),
\end{equation}
where $\sigma_{gas}$ is the surface gas density, $k=1.4$ the law index and $\nu$ the star formation efficiency (SFE). The SFE is assumed to be
$\nu=25$ Gyr$^{-1}$, much higher than what normally assumed in the solar vicinity ($\nu=1$ Gyr$^{-1}$). This is because we assume that the bulge and the most inner Galactic regions suffered a strong starburst.\\
The adopted IMF is the Salpeter (1955) one (with a power index x=1.35). This choice is due to the fact that in Matteucci et al. (2019)  the Calamida et al. (2015) IMF, derived for the Galactic bulge, was also tested and the results did not noticeably differ from those adopting the Salpeter one (see Figure 6 in that paper). In particular, the predicted MDF obtained with Calamida et al. (2015) IMF is slightly shifted towards higher metallicities, with a peak at [Fe/H]$\sim-0.1$ dex, to be compared with the peak predicted by the Salpeter IMF occurring at [Fe/H]$\sim -0.15$ dex. Moreover, the results with Salpeter IMF better agree with the observed distribution for [Fe/H] $> +0.5$ dex.

\subsection{Stellar Nucleosynthesis}
Concerning stellar nucleosynthesis, we adopt as reference yields those of
Romano et al. (2010) which best reproduce the abundance patterns in the solar neighbourhood (Model 15 in that paper).
In particular, the yields of metals from massive stars ($M\ge 10M_{\odot}$) except those of C, N, O, are taken from Kobayashi et al. (2006) and include mass loss depending on metallicity. These stars end their lives as CC-SNe: we adopt the set of yields assuming that a fraction 0.5 of all stars with M$>$20$M_{\odot}$ end up as hypernovae.  In that paper, the mass-cut of the Fe core, which determines the amount of ejected mass relative to that remaining in the neutron star is fixed in order to obtain always 0.07$M_{\odot}$ of ejected Fe, independently of the initial stellar mass. The C, N, O yields from massive stars are instead those computed by the Geneva Group (Meynet \& Maeder, 2002; Hirschi et al. 2005; Hirschi, 2007; Ekstr\"om et al. 2008) including mass loss and rotation and depending also on metallicity.\\
Finally, for the SNe Ia, supposed to originate from white dwarfs in binary systems, the yields are constant with metallicity and relative to the solar chemical composition. These yields, in fact, depend negligibly on the original chemical composition of the stars originating the exploding white dwarf. The assumed yields are from Iwamoto et al. (1999).
The progenitor model for Type Ia SNe is basically the single degenerate one adopted in all previous models for the bulge of Matteucci and collaborators. This progenitor model produces results very similar to the double-degenrrate one, as shown in Matteucci et al. (2009).\\
Concerning low and intermediate mass stars (LIMS, $0.8 \le M/M_{\odot} \le 8$), although they do not contribute to the metals studied here, we adopt the yields of Karakas (2010) up to 6$M_{\odot}$ and we interpolate the yields from 6 to 13$M_{\odot}$ (see Romano et al. 2010).\\
In our models, we keep fixed the IMF, the SFE and the time scale of the infall, $\tau$,  but vary the stellar yields. In particular, we will create empirical yields able to reproduce at best the observed abundance patterns.

\section{Results}
Here, we present the comparisons between our model predictions and the data of Zasowski et al. (2019), and for O and Mg also with the data of Johnson et al. (2014). After these comparisons, we suggest the possible modifications to the adopted stellar yields in order to fit the observed abundance patterns. The yields corrections are obtained by eye fitting the overall distribution in [Fe/H] of each individual chemical abundance with the model. This ensures that we get the best fit for each element over the whole metallicity range.\\
The time-delay model for chemical enrichment (Matteucci, 2012) allows us to interpret the diagrams [X/Fe] vs. [Fe/H] for any chemical element. In fact, since CC- SNe (massive stars) produce the bulk of $\alpha$-elements and only a small fraction of Fe and Fe-peak elements on very short timescales, whereas SNe Ia contribute to the bulk of Fe on longer timescales (exploding white dwarfs), which can be as long as a Hubble time, the [$\alpha$/Fe] ratios are oversolar in the early galaxy evolutionary phases (e.g. at very low metallicities).  Then when SNe Ia start restoring the bulk of Fe, the [$\alpha$/Fe] ratio decreases down to the solar value and below. Let us take the solar vicinity region as a reference: if the star formation rate is higher than in the solar vicinity, the higher than solar [$\alpha$/Fe] ratios would extend for a larger range of [Fe/H] than in the solar region.
In particular, if the ``knee'' where the slope of the [$\alpha$/Fe] ratio changes, occurs at roughly [Fe/H]=-1.0 dex in the solar vicinity, in the bulge, where the SFR has been higher, it should occur at a higher [Fe/H], as predicted by Matteucci \& Brocato (1990) who suggested a knee at [Fe/H] $\sim$ 0.0 dex.

In the Zasowski et al. (2019) data there is a slight decrease with galactocentric distance of the [Fe/H] value at which the knee occurs, but they considered only the inner 4 kpc where we should not expect a sensitive variation of the plateau of the $\alpha$-enhancement.  The reason for having a longer plateau in the bulge than in the solar vicinity, is that the intense bulge SFR creates many CC- SNe which enrich the interstellar medium (ISM) with Fe (although they are not the major producers of this element), and when the SNe Ia start occurring (the time is fixed by stellar evolution) the ISM [Fe/H] is higher than for a lower SFR regime. The opposite occurs if the SFR is lower than the one in the solar vicinity.

For irregular dwarf galaxies, in fact, we expect a knee at lower [Fe/H] values than in the solar neighbourhood and low [$\alpha$/Fe] ratios at low metallicity. This fact is observed in dwarf spheroidal galaxies (e.g. Lanfranchi \& Matteucci 2003, Tolstoy et al. 2009).
It is not very easy to establish where the knee occurs in the bulge relative to the solar vicinty, because of the very high number of stars available at the present time and the consequent spread observed in the abundance ratios.
If an element is produced mainly by long living stars, and partly by massive stars, such as Fe, whose bulk is produced by exploding C-O white dwarfs, then the [el/Fe] ratio should be constant all over the [Fe/H] range.
Finally, if an element is produced in a secondary fashion, namely proportionally to the abundance of metals present in the star since its birth, then the [X/Fe] should increase with [Fe/H]

In the following, we will interpret our result according to the paradigm of the time-delay model.

\subsection{The $\alpha$-elements: O, Mg, Si, Ca}

The so called $\alpha$-elements are those produced by fusion of $\alpha$-particles and they are produced during both hydrostatic and explosive burnings in massive stars. They are O, Ne, Mg, Si, Ca. Before discussing element by element, we remind here that Prantzos et al. (2018) tested yields of $\alpha$-elements with and without stellar rotation and concluded that rotation does not affect them. Our reference yields do not include rotation except for oxygen. The reason why we adopt O yields with stellar rotation is only because we are adopting the best set of yields suggested in the paper of Romano et al. (2010), who found that to reproduce the evolution of C, N and O in the solar vicinity, the yields with rotation of the Geneva Group should be preferred.

\subsubsection{Oxygen}

\begin{figure*}
  \includegraphics[width=1.15\columnwidth]{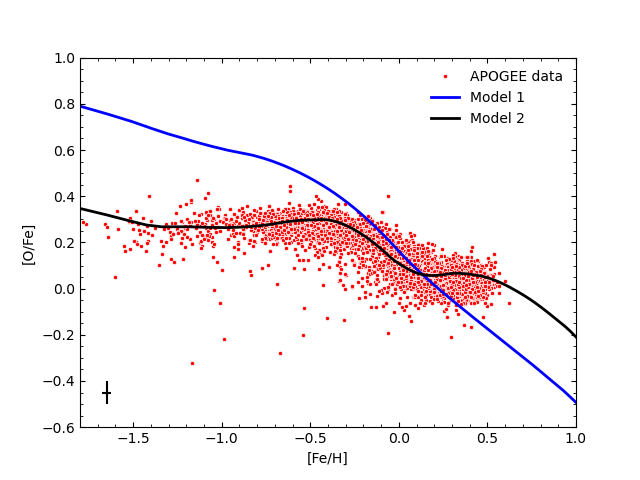}
 \includegraphics[width=1.15\columnwidth]{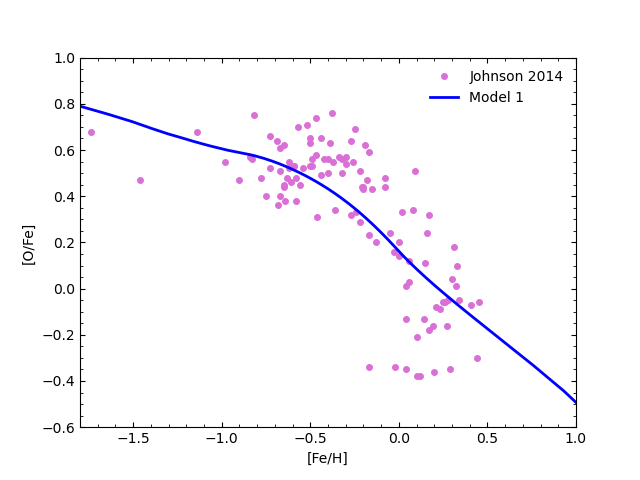}
    \caption{Predicted and observed [O/Fe] vs. [Fe/H] in the bulge region. Upper panel: the blue curve labelled Model 1 is our reference model adopting stellar yields from the literature (see text), while the black curve represents Model 2, namely the model obtained by correcting the stellar yields, as described in the text. The data are from Zasowski et al. (2019) and the error bars are indicated in the lower left corner of the Figure. Lower panel: Model 1 results compared to the bulge data of Johnson et al. (2014).}
    \label{fig1}
\end{figure*}
In Figure 1, we show the [O/Fe] vs. [Fe/H] diagram. We refer to the $^{16}$O isotope which is the third most abundant element in the Universe, after H and He. Oxyen is originated during He-burning by means of the reaction
$^{12}$C($\alpha, \gamma$)$^{16}$O and also by photodisintegration of $^{20}$Ne in massive stars ($M \ge 10M_{\odot}$) ending as CC-SNe. The yield of O increases with the stellar mass. In the upper panel of Figure 1, the data are the red points while the models are represented by the continuous lines. The blue line represents Model 1, namely the standard model obtained with the assumed set of yields, as described in the previous Section. As one can see, the blue line does not fit well the data: in particular, it predicts a too high [O/Fe] ratio for metallicity below [Fe/H]=-1.0 dex, and a too steep decline for [Fe/H]$ \ge$ 0.0 dex.\\
The black line in Figure 1 represents Model 2, namely the model with corrected O yields, and it fits quite nicely the data. It should be noted, that previous data, such as those of Johnson et al. (2014), are in agreement with Model 1, as shown in the lower panel of Figure 1. This discrepancy could be due in part to the fact that the infrared lines are used to derive the O abundance, and they lie in convective regions. Thus corrections for convection, computed by means of atmosphere models, should be applied to these abundance determinations. Therefore, no firm conclusions can be drawn on the oxygen.

From the theoretical point of view, the behaviour of the [O/Fe] ratio is interpreted in the framework of the time-delay model, and arises from the fact that O is mainly produced by massive stars, and therefore [O/Fe] is almost constant at low metallicities and then declines toward the solar value as the metallicity increases. The flattening of the same ratio at  metallicity larger than solar is not clear and probably not real.

Another fact to consider is that there is a small group of stars in Zasowski et al. (2019) data with [Fe/H]=0 and [O/Fe]=+0.3 dex detaching themselves significantly from the other stars of the sample. It is possible that these stars can be affected by some error in the abundance derivation process. In fact, the feature in O has been reanalysed by ASPCAP, and  J\"onsson et al.
(2020, submitted) concluded that  this feature is due to some possible
systematics in the abundance determinations, which the APOGEE team has
not yet been able to identify.

In order to obtain Model 2 (black line) providing a good fit to the data, the yields of O had to be modified in the following way: we had to change the O yields of massive stars by different factors according to the initial stellar metallicity, in particular for low metallicities the standard yields were lowered by a factor 0.45, and those relative to high metallicities were increased by a factor 3.5 (see Table 1). \\

\subsubsection{Magnesium}
  In Figure 2,  we present the [Mg/Fe] vs. [Fe/H] diagram. We refer to the isotope $^{24}$Mg which is formed during shell C-burning in massive stars. The standard model is again indicated by Model 1 and the corrected one by Model 2. Also here the standard model predicts a too high overabundance of Mg relative to Fe at low metallicities. Therefore, also in this case we had to lower the Mg yields of massive star relative to low metallicities by a factor 0.65 leaving the yields of more metal rich stars untouched. Similar results were obtained for the bulge by Matteucci et al. (2019) who considered APOGEE data and reached similar conclusions. This discrepancy is not present when the model is compared to Gaia-ESO data and the yields of Kobayashi et al. (2006) fit well the [Mg/Fe] (see Matteucci et al. 2019) as well as the data of Johnson et al. (2014), shown also in Figure 2 (lower panel). The difference between APOGEE, Gaia-ESO and Johnson et al. (2014) data can be due to different calibrations adopted in the data reduction process in the different data samples. On the other hand, the model does not reproduce the flattening of [Mg/Fe] observed at  high metallicity and in Matteucci et al. (2019) it was suggested that it can be obtained by assuming increased Mg yields from SNe Ia. The increase should be as high as a factor of 10, but this is probably a non realistic suggestion on the basis of the nucleosynthesis calculations for SNe Ia (Iwamoto et al.1999).  However, Gaia-ESO as well as  Johnson's et al. (2014) data do not show this flattening and our Mg curve fits them well (see Figure 2 lower panel), therefore it is premature to draw conclusions on the Mg behaviour at high metallicities, if different data sets are not in agreement. So, we conclude like for oxygen that this flattening is probably an artefact.\\

  \begin{figure*}
    \includegraphics[width=1.15\columnwidth]{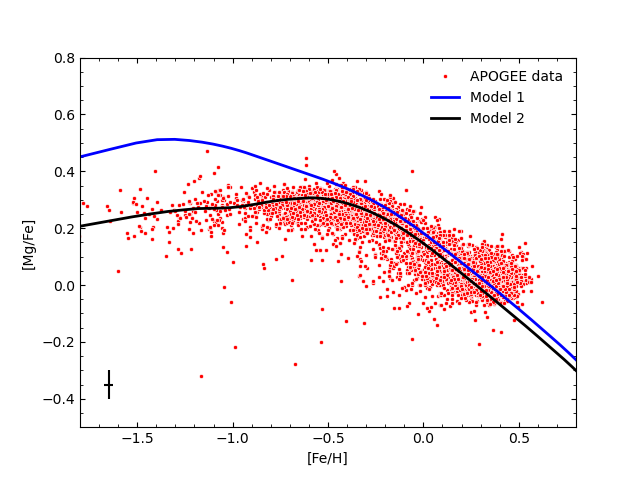}
    \includegraphics[width=1.15\columnwidth]{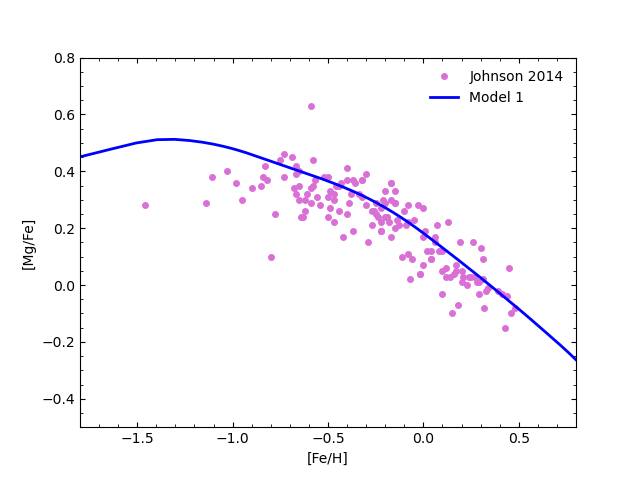}
    \caption{Predicted and observed [Mg/Fe] vs. [Fe/H] in the bulge region. Upper panel: the lines referring to the models represent the standard one (blue line) and the corrected one (black line), the data are from Zasowski et al. (2019) and the error bars are indicated in the lower left corner of the Figure, as in Figure 1. Lower panel:The results of Model1 compared to the bulge data of Johnson et al. (2014).}
	    \label{fig2}
\end{figure*}

\subsubsection{Silicon}
In Figure 3, we show [Si/Fe] vs. [Fe/H]. The isotope we refer is $^{28}$Si which originates from carbon and neon burning processes, both explosive and hydrostatic, and is produced both in CC-SNe and Type Ia SNe in almost equal proportions, if we integrate the yields on the IMF (see Matteucci 2001). Here, the standard model reproduces the trend but it lies at too high [Si/Fe] ratios over the whole [Fe/H] range, so to obtain Model 2 we simply lowered all the yields from massive stars by a factor 0.6 independently of the stellar metallicity. At high metallicities also this element shows a slight flattening which is probably not real. The data of Johnson et al. (2014), in fact,  do not show any flattening for this element.\\

\begin{figure}
		\includegraphics[width=1.15\columnwidth]{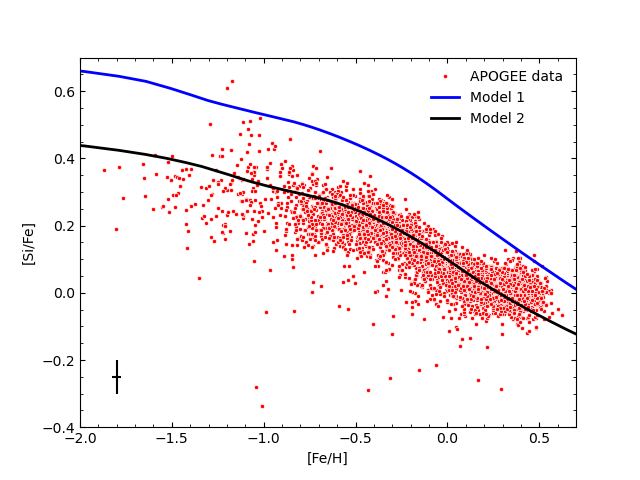}
    \caption{Predicted and observed [Si/Fe] vs. [Fe/H] in the bulge region. The lines referring to the models represent the standard one (blue line) and the corrected one (black line), the data are from Zasowski et al. (2019) and the error bars are indicated in the lower left corner of the Figure as in Figure 1.}
    \label{fig4}
\end{figure}

\subsubsection{Calcium}
In Figure 4, the [Ca/Fe] vs. [Fe/H] diagram is presented. We refer to the isotope $^{40}$Ca formed during C- and O-burning processes (both hydrostatic and explosive). It can be produced both in massive stars and in SNe Ia, almost in equal proportions (the same situation as Si). Here,  Model 1 predicts too low values of [Ca/Fe] for [Fe/H] $>0.0$ dex. Again a trend of flat [Ca/Fe] is found at high metallicity as for the other $\alpha$-elements.\\
In order to fit the data and obtain Model, 2 we  had to correct only  the yields from Type Ia SNe which contribute to chemical enrichment with a time delay and therefore at high metallicity. We increased the yield of Ca from SNe Ia by introducing a dependence of it on metallicity.
This allows us to reproduce the plateau observed at high metallicity for the  [Ca/Fe] ratio. Again, other data such as those of Johnson et al. (2014) do not show such a plateau. Therefore, even in this case the plateau is probably an artefact.
\\

\begin{figure}
		\includegraphics[width=1.15\columnwidth]{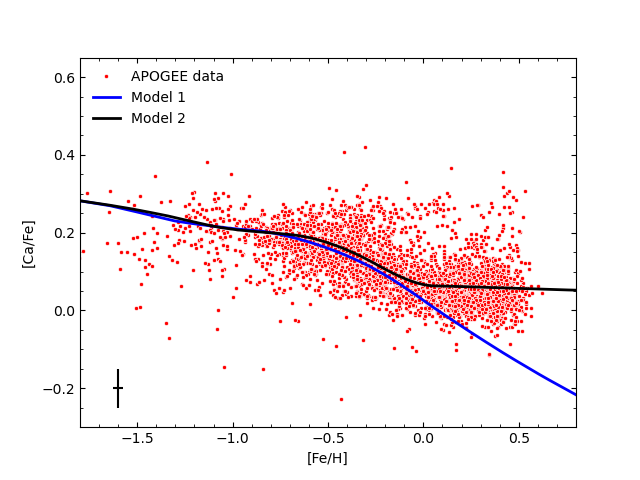}
                \caption{Predicted and observed [Ca/Fe] vs. [Fe/H] in the bulge region. The lines referring to the models represent the standard one (blue line) and the corrected one (black line), the data are from Zasowski et al. (2019) and the error bars are indicated in the lower left corner of the Figure, as in Figure 1.}
  \label{fig5}
  \end{figure}

\subsection{The odd elements: Al, K}
The odd-elements possess an odd number of protons.

\subsubsection{Aluminum}
In Figure 5, we present [Al/Fe] vs. [Fe/H]. We refer to the isotope $^{27}$Al, since $^{26}$Al is a radioactive element, and is produced during C-burning and explosive Ne burning in massive stars. In C-burning, two C atoms fuse to give rise to Mg isotopes. In particular, once $^{26}$Mg is formed, its reactions with free protons and neutrons synthesize $^{27}$Al. At a first sight, $^{27}$Al seems to be a primary element originating from H and He from which C is formed, but it has been shown that $^{27}$Al should have a secondary behaviour since its abundance depends on the amount of $^{22}$Ne burned during C-burning and  $^{22}$Ne depends on the abundances of C and O originally present in the star (Clayton, 2007). The data here show a large dispersion but also a secondary behaviour, at least at low metallicities. This element should be produced in a negligible way by Type Ia SNe. Here, Model 1 is simply shifted at higher [Al/Fe] ratios than the data over the whole range of [Fe/H]. Therefore, to obtain Model 2 we had to simply lower the yields of massive stars by a constant 0.8 factor. It should be noted that other studies like that of Johnson et al. (2014) show a [Al/Fe] behaviour very similar to that of $\alpha$-elements, whereas in Alves-Brito et al. (2010), Fulbright et al. (2007) and Lecureur et al. (2007) there is a behaviour similar to the one of Figure 5.
\begin{figure}
		\includegraphics[width=1.15\columnwidth]{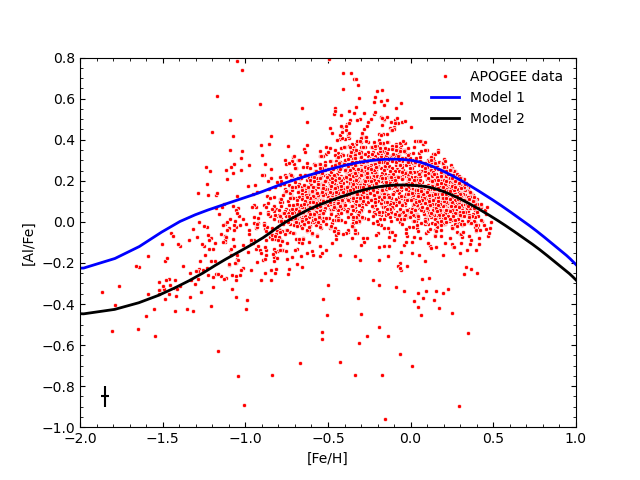}
    \caption{Predicted and observed [Al/Fe] vs. [Fe/H] in the bulge region. The lines referring to the models represent the standard one (blue line) and the corrected one (black line), the data are from Zasowski et al. (2019) and the error bars are indicated in the lower left corner of the Figure, as in Figure 1.}
    \label{fig6}
\end{figure}

\subsubsection{Potassium}
In Figure 6, we show [K/Fe] vs. [Fe/H]. We refer to the isotope $^{41}$K which originates from the O-burning in massive star explosions.
The nucleosynthesis situation of K is complicated by the fact that it can originate also in neutrino-induced reactions. It is first produced as $^{41}$Ca which decays into $^{41}$K.
This element appears peculiar, since in principle it should behave as an $\alpha$ -element,  being produced mainly by massive stars. The data show instead an almost constant plateau with [K/Fe]$\sim$ +0.1 dex followed by a decrease to the solar value starting at [Fe/H]$\sim$ -0.5 dex and  followed by an increase at metallicities larger than solar. Here, Model 1 predicts a much too low value for the [K/Fe] ratio all over the [Fe/H] range. The required corrections are therefore very large. This mismatch between yields and data of K in the solar neighbourhood had been already noted by Fran\c cois et al. (2004), Romano et al. (2010) and Prantzos et al. (2018), who suggested that yields with stellar rotation (such as those of Chieffi \& Limongi, 2003;2004, Limongi \& Chieffi 2018) can improve the K yields at low metallicities. Here, we find the same problem for the inner Galactic regions. \\
To obtain Model 2, the yields of K from massive stars had to be multiplied by a factor of 7 independent of metallicity and then by other minor factors according to the metallicity.\\

\begin{figure}
		\includegraphics[width=1.15\columnwidth]{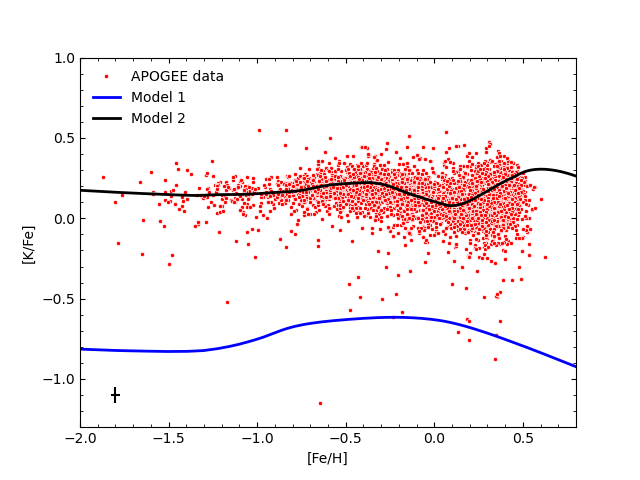}
                \caption{Predicted and observed [K/Fe] vs. [Fe/H] in the bulge region. The lines referring to the models represent the standard one (blue line) and the corrected one (black line), the data are from Zasowski et al. (2019) and the error bars are indicated in the lower left corner of the Figure, as in Figure 1.}
                
    \label{fig7}
\end{figure}

\subsection{The Fe-peak elements: Cr, Mn, Ni}
The Fe-peak elements should be mainly produced by Type Ia SNe, although CC-SNe produce part of them. It is worth noting that the yields from Type Ia SNe in principle do not depend on metallicity.
\subsubsection{Chromium}
In Figure 7 we analyze the [Cr/Fe] vs. [Fe/H]. We refer to $^{52}$Cr the most abundant of Cr isotopes. It is formed  during explosive Si-burning with incomplete Si-exhaustion both in massive stars (CC-SNe) and Type Ia SNe. This element belongs to the Fe-peak elements. It follows roughly the behaviour of Fe, as expected from the time-delay model, at least until [Fe/H]=-0.2 dex. For larger metallicities, the [Cr/Fe] ratio decreases slightly. This decrease is not reproduced by Model 1. Here the required corrections are small, both for the yields of Type Ia SNe and massive stars. In particular, we have multiplied the yields of massive stars by a factor 1.1 and those by SNe Ia by  0.5.\\

\begin{figure}
		\includegraphics[width=1.15\columnwidth]{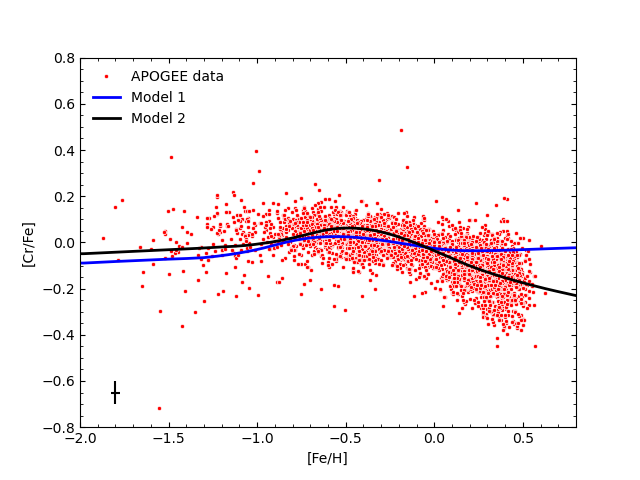}
                \caption{Predicted and observed [Cr/Fe] vs. [Fe/H] in the bulge region. The lines referring to the models represent the standard one (blue line) and the corrected one (black line), the data are from Zasowski et al. (2019) and the error bars are indicated in the lower left corner of the Figure, as in Figure 1.}
                
    \label{fig8}
\end{figure}

\subsubsection{Manganese}
In Figure 8 we find [Mn/Fe] vs. [Fe/H]. We refer to the isotope $^{55}$Mn which is produced during explosive Si-burning with incomplete Si exhaustion and  $\alpha$-rich freeze-out either in massive stars (CC-SNe) or Type Ia SNe. It is the result of the decay of $^{55}$Co. The general trend of [Mn/Fe] is increasing with [Fe/H]. This element is produced in almost equal amounts by the core-collapse SNe and the Type Ia SNe. Model 1 here lies below the observed points. In order to obtain Model 2 we adopted the suggestion of Cescutti et al. (2008) who introduced a multiplicative factor to the yields of Mn of $(Z/Z_{\odot})^{0.65}$ (where $Z$ is the global metallicity) from SNe Ia, plus a multiplicative factor of 1.8 applied to the yields of massive stars. The dependence of the yield of Mn from SNe Ia upon the metallicity seems to be the most important assumption to reproduce the typical secondary behaviour of this element. Cescutti \& Kobayashi (2017) suggested that Mn can be produced by new sub-classes of SNeIa.\\

\begin{table*}
	\centering
	\caption{Summary of corrections to apply to the standard yields.}
	\label{tab:example_table}
	\begin{tabular}{lccr} 
		\hline
		Element & CC-SNe & TypeIa SNe \\
		\hline
		$^{16}O$ & x0.45 for [Fe/H] $<$-1.0, x3.5 for higher [Fe/H] \\
		$^{24}Mg$ & x0.65 for [Fe/H]$<$-1.0 & --\\
		$^{28}Si$  & x0.6 independent of [Fe/H] & -- \\
                $^{40}Ca$  &-----& x3 for  [Fe/H]$>$ 0.0
                \\
                $^{27}Al$ & x0.8 independent of [Fe/H]&---\\
                $^{41}K$  & x7 independent of [Fe/H] & --\\
                $^{52}Cr$  & x1.1 independent of [Fe/H] & x0.5 \\
                $^{55}Mn$  & x1.8 independent of [Fe/H] & yields going as $(Z/Z_{\odot})^{0.65}$
               \\
                $^{58}Ni$  & x 1.5 independent of [Fe/H] & x0.15 &\\
		\hline
	\end{tabular}
\end{table*}

\begin{figure}
		\includegraphics[width=1.15\columnwidth]{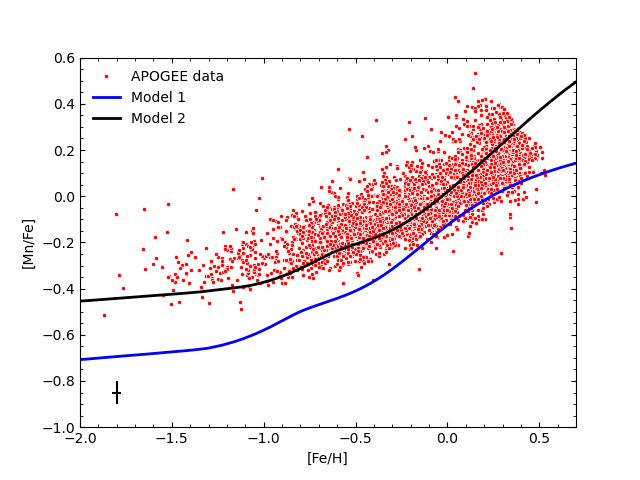}
                \caption{Predicted and observed [Mn/Fe] vs. [Fe/H] in the bulge region. The lines referring to the models represent the standard one (blue line) and the corrected one (black line), the data are from Zasowski et al. (2019) and the error bars are indicated in the lower left corner of the Figure, as in Figure 1.}
                        \label{fig9}
\end{figure}

\subsubsection{Nickel}
In Figure 9, we show the plot [Ni/Fe] vs. [Fe/H]. We refer to the isotope $^{58}$Ni which is the most abundant stable isotope of Ni. In fact, $^{56}$Ni and $^{57}$Ni are radioactive nuclei. It is produced in explosive Si-burning with complete Si exhaustion. Again, Ni is a Fe-peak element and it should be produced more by SNe Ia than CC-SNe. Here, Model 1 clearly does not reproduce the data behaviour which is almost flat until [Fe/H]=0.0 dex and then it increases slightly. The large increase of [Ni/Fe] with metallicity of Model 1 is due to the large yield of Ni in Type Ia SNe (Iwamoto et al. 1999).\\
To obtain Model 2 we had to increase the yields of massive stars as functions of metallicity by factors between 1.4 and 1.6 (average 1.5), and to decrease the Ni production by SNe Ia by an average factor of 0.15 which increases with metallicity. Metallicity dependent yields for Ni from SNe Ia had already been suggested by Zasowski et al. (2019) to reproduce the small increase of [Ni/Fe] for [Fe/H]$>$ 0.0 dex.\\ 

\begin{figure}
		\includegraphics[width=1.15\columnwidth]{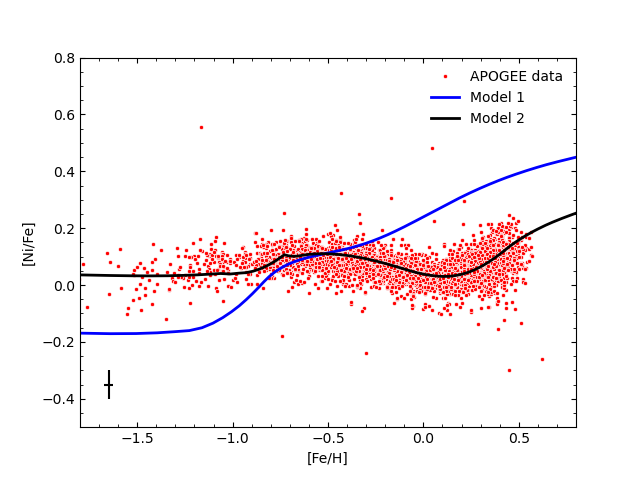}
                \caption{Predicted and observed [Ni/Fe] vs. [Fe/H] in the bulge region. The lines referring to the models represent the standard one (blue line) and the corrected one (black line), the data are from Zasowski et al. (2019) and the error bars are indicated in the lower left corner of the Figure, as in Figure 1.} 
                
    \label{fig11}
\end{figure}

\section{Discussion and Conclusions}

In this paper, we have computed the chemical evolution of the inner Galactic regions by adopting a successful model for the Galactic bulge, assuming a strong and short star formation burst which originated most of the stars. We computed the evolution of the abundances of 11 chemical species (O, Mg, Al, Si, K, Ca, Cr, Mn, Fe, Ni) and compared the results with the data of Zasowski et al. (2019) including roughly 4000 stars.
We adopted a set of stellar yields from core-collapse and Type Ia SNe already tested on the abundance patterns observed in the solar neighbourhood, and found that in order to obtain a good fit to the data the yields of some elements need to be corrected. These corrections are indeed artificial but suggest how to modify
some physical inputs in stellar nucleosynthesis models which contain uncertainties relative to the rates of important nuclear reactions, as well as mechanisms of explosion of SNe, mass cut in the collapsing Fe core before the explosion, stellar rotation, mass loss and treatment of convection. It is worth noting that since the uncertainties in the data are $<0.1$dex and the required yield variations produce results which differ by more than that, we conclude that the observational uncertainties do not affect the adjusted yields.\\
The adopted data show some peculiar behaviour especially for the $\alpha$-elements. In particular, they show a flattening of the [$\alpha$/Fe] ratio for [Fe/H]$>$0.0 dex not visible in other data sets, such as the Gaia-ESO and Johnson et al. (2014) ones. This fact, if real, requires strong variations (increase) of the yields of the $\alpha$-elements from Type Ia SNe which instead are believed to produce mostly Fe and Fe-peak elements. May be, this flattening could be an artifact due to differences in the calibration among different data sets. Therefore, without entering into details of the procedures of data reduction, we simply suggest that no firm conclusions can be drawn on this  subject. The suggested yields variations are summarized in Table 1.\\

Our more detailed suggestions can be summarized as follows:
\begin{itemize}
\item For Si, Ca, Cr and Ni, the required corrections are small. These elements are produced both in core-collapse and Type Ia SNe. In particular, for Si it is enough to lower the yields from massive stars by a factor 0.6 independently of metallicity. For Ca, it is enough to increase the yields from SNe Ia but introducing a metallicity dependence. For Cr and Ni, small corrections are required for both the yields from Type Ia and core-collapse SNe.

  \item The corrections required for Mg and Al are moderate. Both these elements are mainly produced by massive stars. The corrections required for Mg were already discussed in Matteucci et al. (2019) and consist in lowering the yield of massive stars at low metallicity by a factor of 0.65. For Al, a constant correction factor of 0.8 needs to be applied to the yields from massive stars, independently of metallicity.

  \item The corrections required for O and K are the largest and more complex ones. The same problem arises for K in the solar vicinity, as pointed out by Romano et al. (2010). For oxygen, the situation appears unusual since the standard yields are able to well reproduce the [O/Fe] vs. [Fe/H] trend in the solar vicinity (Romano et al. 2010).
 For potassium, the standard yields do more or less reproduce the trend [K/Fe] vs. [Fe/H], but largely underestimate the absolute values of the [K/Fe] ratio at all metallicities. In order to reproduce the data, we had to increase the standard yields from massive stars  by a factor of 7, plus smaller corrective factors depending on metallicity to improve the trend.

  \item Finally, the corrections required for Mn are not new (see Cescutti et al. 2008). The behaviour of this element is a secondary one and this is well reproduced if the yields from Type Ia SNe are metallicity dependent.

 \end{itemize}

\section*{Acknowledgements}
F.M. thanks M. Limongi for many useful suggestions on stellar nucleosynthesis. We thank an anonymous referee for her/his careful reading of the paper and useful suggestions that improved the paper.





\begin{thebibliography}{99}

\bibitem[\protect\citeauthoryear{}{}]{b23}
Abolfathi, B. et al., 2018, ApJS, 235, 42

  
\bibitem[\protect\citeauthoryear{}{}]{b23}
Alves-Brito A., Mel{\'e}ndez J., Asplund M., Ram{\'\i}rez I., Yong D., 2010, A\&A, 513, A35

\bibitem[\protect\citeauthoryear{}{}]{b23}
Ballero, S.K., Matteucci, F., Origlia, L., \& Rich, R.M.\ 2007, \aap, 467, 123 

\bibitem[\protect\citeauthoryear{}{}]{b23}
Bensby T., et al., 2013, A\&A, 549, A147

\bibitem[\protect\citeauthoryear{}{}]{b23}
Bensby T., et al., 2017, A\&A, 605, A89

\bibitem[\protect\citeauthoryear{}{}]{b23}
  Calamida, A., Sahu, K.~C., Casertano, S., et al.\ 2015, \apj, 810, 8
  
\bibitem[\protect\citeauthoryear{}{}]{b23}
  Cescutti, G. \& Kobayashi, C., 2017, A\&A, 607, 23

\bibitem[\protect\citeauthoryear{}{}]{b23}
Cescutti G., Matteucci F., Lanfranchi G.~A., McWilliam A., 2008, A\&A, 491, 401

\bibitem[\protect\citeauthoryear{}{}]{b23}
Cescutti, G., \& Matteucci, F. 2011, \aap, 525, A126 

\bibitem[\protect\citeauthoryear{}{}]{b23}
Chieffi A., Limongi M., 2003, PASA, 20, 324
  
\bibitem[\protect\citeauthoryear{}{}]{b23}
Chieffi A., Limongi M., 2004, ApJ, 608, 405
  
\bibitem[\protect\citeauthoryear{}{}]{b23}
Clarkson W., et al., 2008, ApJ, 684, 1110
  
\bibitem[\protect\citeauthoryear{}{}]{b23}
Clayton D., 2007, Handbook of Isotopes in the Cosmos, Cambridge University Press

\bibitem[\protect\citeauthoryear{}{}]{b23}
  C{\^o}t{\'e} B., O'Shea B.~W., Ritter C., Herwig F., Venn K.~A., 2017, ApJ, 835, 128
  
\bibitem[\protect\citeauthoryear{}{}]{b23}
Cunha, K. \& Smith, V.V, 2006, ApJ, 651, 491
  
\bibitem[\protect\citeauthoryear{}{}]{b23}
Ekstr{\"o}m S., Meynet G., Chiappini C., Hirschi R., Maeder A., 2008, A\&A, 489, 685

\bibitem[\protect\citeauthoryear{}{}]{b23}
Fran{\c c}ois, P., Matteucci, F., Cayrel, R., et al. 2004, A\&A, 421, 613
  
\bibitem[\protect\citeauthoryear{}{}]{b23}
Fulbright J.~P., McWilliam A., Rich R.~M., 2007, ApJ, 661, 1152

\bibitem[\protect\citeauthoryear{}{}]{b23}
Garcia P\'erez, A.E. et al., 2016, AJ, 151, 144

\bibitem[\protect\citeauthoryear{}{}]{b23}
Gonzalez O.~A., et al., 2015, A\&A, 584, A46

\bibitem[\protect\citeauthoryear{}{}]{b23}
Grieco, V., Matteucci, F., Pipino, A., \& Cescutti, G., 2012, \aap, 548, A60 

\bibitem[\protect\citeauthoryear{}{}]{b23}
Iwamoto K., Brachwitz F., Nomoto K., Kishimoto N., Umeda H., Hix W.~R., Thielemann F.-K., 1999, ApJS, 125, 439

  
\bibitem[\protect\citeauthoryear{}{}]{b23}
Hirschi R., Meynet G., Maeder A., 2005, A\&A, 433, 1013

\bibitem[\protect\citeauthoryear{}{}]{b23}
Hirschi R., 2007, A\&A, 461, 571
  
\bibitem[\protect\citeauthoryear{}{}]{b23}
Johnson C.~I., Rich R.~M., Kobayashi C., Kunder A., Koch A., 2014, AJ, 148, 67

\bibitem[\protect\citeauthoryear{}{}]{b23}
  J{\"o}nsson H., et al., 2018, AJ, 156, 126
  
\bibitem[\protect\citeauthoryear{}{}]{b23}
J{\"o}nsson H., et al., 2020, submitted

\bibitem[\protect\citeauthoryear{}{}]{b23}
Lanfranchi G.~A. \& Matteucci F., 2003, MNRAS, 345, 71

\bibitem[\protect\citeauthoryear{}{}]{b23}
Lecureur A., et al., 2007, A\&A, 465, 799

\bibitem[\protect\citeauthoryear{}{}]{b23}
Limongi M., Chieffi A., 2018, ApJS, 237, 13

\bibitem[\protect\citeauthoryear{}{}]{b23}
Karakas A. I., 2010, MNRAS, 403, 1413

\bibitem[\protect\citeauthoryear{}{}]{b23}
Kennicutt, R. C., Jr, 1998, ApJ, 498, 541

\bibitem[\protect\citeauthoryear{}{}]{b23}
Kobayashi C., Umeda H., Nomoto K., Tominaga N., Ohkubo T., 2006, ApJ, 653, 1145

\bibitem[\protect\citeauthoryear{}{}]{b23}
Majewski S.~R., et al., 2017, AJ, 154, 94

\bibitem[\protect\citeauthoryear{}{}]{b23}
Matteucci, F. 2001, The Chemical Evolution of the Galaxy, ASSL, Kluwer
Academic Publisher

\bibitem[\protect\citeauthoryear{}{}]{b23}
Matteucci F., 2012, Chemical Evolution of Galaxies. Springer-Verlag, Berlin

\bibitem[\protect\citeauthoryear{}{}]{b23}
Matteucci, F., \& Brocato, E.\ 1990, \apj, 365, 539
  
\bibitem[\protect\citeauthoryear{}{}]{b23}
Matteucci F., Spitoni E., Recchi S., Valiante R., 2009, A\&A, 501, 531
  
\bibitem[\protect\citeauthoryear{}{}]{b23}
Matteucci F., Grisoni V., Spitoni E., Zulianello A., Rojas-Arriagada A., Schultheis M., Ryde N., 2019, MNRAS, 487, 5363

\bibitem[\protect\citeauthoryear{}{}]{b23}
McWilliam, A. \& Rich, R.M., 1994, ApJS, 91, 749

\bibitem[\protect\citeauthoryear{}{}]{b23}
Meynet G., Maeder A., 2002, A\&A, 390, 561

\bibitem[\protect\citeauthoryear{}{}]{b23}
Nidever, D.L., 2015, AJ, 150, 173

\bibitem[\protect\citeauthoryear{}{}]{b23}
Prantzos N., Abia C., Limongi M., Chieffi A., Cristallo S., 2018, MNRAS, 476, 3432

\bibitem[\protect\citeauthoryear{}{}]{b23}
Rojas-Arriagada A., et al., 2017, A\&A, 601, 140
  
\bibitem[\protect\citeauthoryear{}{}]{b23}
Romano D., Karakas A. I., Tosi M., Matteucci F., 2010, A\&A, 522, A32

\bibitem[\protect\citeauthoryear{}{}]{b23}
Salpeter, E.~E.\ 1955, \apj, 121, 161 

\bibitem[\protect\citeauthoryear{}{}]{b23}
Tolstoy, E., Hill, V., Tosi, M., 2009, ARA\&A, 43, 371

\bibitem[\protect\citeauthoryear{}{}]{b23}
Woosley S.~E., Weaver T.~A., 1995, ApJS, 101, 181

\bibitem[\protect\citeauthoryear{}{}]{b23}
  Zasowski G., et al., 2019, ApJ, 870, 138
  
  \bibitem[\protect\citeauthoryear{}{}]{b23}
Zasowski G., et al., 2013, AJ, 146, 81
  
\bibitem[\protect\citeauthoryear{}{}]{b23}
Zoccali M., et al., 2003, A\&A, 399, 931

\bibitem[\protect\citeauthoryear{}{}]{b23}
Zoccali M., et al., 2006, A\&A, 457, L1



  
\end{thebibliography}







\bsp	
\label{lastpage}
\end{document}